# Si-rich silicon-nitride waveguides for optical transmissions and towards wavelength conversion around 2 μm

**Manon Lamy,**[1] **Christophe Finot,**[1,*] **Alexandre Parriaux,**[1] **Cosimo Lacava,**[2] **Thalia Dominguez Bucio,**[2] **Frederic Gardes,**[2] **Guy Millot,**[1] **Periklis Petropoulos,**[2] **and Kamal Hammani**[1]

[1]*Laboratoire Interdisciplinaire Carnot de Bourgogne, UMR 6303 CNRS-Université Bourgogne-Franche-Comté, 9 av. A. Savary, 21078 Dijon cedex, France*
[2]*Optoelectronics Research Centre, University of Southampton, SO17 1BJ, Southampton, United Kingdom*
*\*christophe.finot@u-bourgogne.fr*

**Abstract:** We show that subwavelength silicon-rich nitride waveguides efficiently sustain high-speed transmissions at 2 μm. We report the transmission of a 10 Gbit/s signal over 3.5 cm with negligible power penalty. Parametric conversion in the pulsed pump regime is also demonstrated using the same waveguide structure with an efficiency as high as - 18 dB.

## 1. Introduction

Internet traffic has experienced exponential growth in recent years and involves a huge amount of telecommunication data that needs to be processed in the optical layer. Integrated photonic devices are crucial for this task. Since the conventional O- or C- bands are already heavily used, a new waveband centered at 2 μm has been proposed, taking advantage of the broad and high gain provided by thulium doped fiber amplifiers (TDFA), as well as the development of hollow core photonic bandgap fibers designed to exhibit a minimum loss around 2 μm [1, 2].This waveband has also attracted the interest of the integrated photonics community, and several components have recently been developed, such as InP-based modulators [3] or arrayed waveguide gratings [4].

In addition to the development of new components, a number of works have emerged in recent years that experimentally explore the potential of various materials in this new spectral range, with an emphasis to components having subwavelength transverse dimensions to ensure a high density of photonic circuitry. For instance, efficient data transmission at 10 Gbit/s in subwavelength components has been reported using different materials such as silicon germanium [5], titanium dioxide [6] or silicon on insulator [7]. Another CMOS-compatible platform that is currently stimulating much research is silicon nitride ($Si_3N_4$) and its non-stoichiometric compounds [8, 9]. However, no high bit-rate transmission has been reported so far at wavelengths around 2 μm for this material. Here, we demonstrate, penalty-free transmission of a 10-Gbit/s on-off keying signal at 1.98 μm in a 3.5-cm long Si-rich nitride waveguide.

We also discuss the possibility to achieve nonlinear processing under pulsed-pump operation. All-optical wavelength conversion allows flexible use of network resources and may even facilitate interoperability between spectral bands. Previous examples have shown the efficiency of a parametric process in tellurite [10] or tapered chalcogenide microstructured fibers around 2 μm [11, 12]. Following the trend of device miniaturization and thanks to their large third-order nonlinear coefficient, silicon-based integrated optical devices have also been considered for the implementation of several ultrafast signal processing functionalities [13, 14].

Experimental demonstrations in silicon around 2 µm have also been reported with either picosecond [15, 16] or continuous wave pumping [7, 17-19]. Silicon nitride is a silicon-based compound that is currently the subject of many efforts [20, 21] thanks to a very low level of optical losses and the absence of two-photon absorption in the near infrared. To overcome the limited nonlinearity of stoichiometric silicon nitride, Si-rich silicon nitride (SRN) has been studied [22, 23] and wavelength conversion [24-26] as well as spectral broadening [27, 28] have been successfully demonstrated in the C-band. Here, we show that the same structure as the one used in the transmission experiments is also suitable for parametric wavelength conversion in the 2-µm range, and reveal a conversion efficiency up to - 20 dB for pulsed operation.

Our article is organized as follows. We first present the design and fabrication of the component with an emphasis on the modal properties of the waveguide. We then cover the setups used for the experimental demonstrations carried out around 2 µm. After having established that error-free high-speed transmission can be achieved, we end by discussing the wavelength conversion observed in these waveguides.

## 2. Waveguide design and fabrication

The compact integrated planar waveguides used in this work are fabricated based on a rigorous development campaign, detailed in [29] and [30]. A schematic of the waveguide cross section is shown in Fig. 1(a). A 300 nm thick layer of silicon-rich silicon nitride was deposited on a thermally oxidized silicon wafer with a $SiO_2$ thickness of 2 µm. The Inductive Coupled Plasma (ICP) etching technique was used to pattern the waveguides into the silicon-rich silicon nitride layer, followed by the deposition of a protective 1 µm-thick $SiO_2$ cladding layer. The photonic chip included a set of 3 cm- long waveguides with various widths ranging from 1.4 µm up to 2.2 µm.

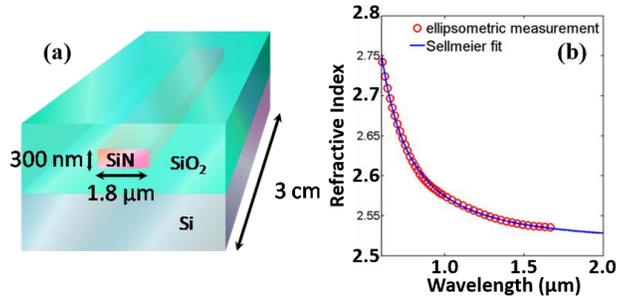

Fig. 1. (a) Structure of the waveguides under investigation (1.8 µm width).(b) Ellipsometric recording and Sellmeier fit.

Use of a commercial finite difference eigenmode solver enabled us to simulate the modal content of the waveguide. We assumed an index value of $n_{SiO2} = 1.438$ for the $SiO_2$ layer and $n_{SRN} = 2.528$ for the SRN at 2 µm in these simulations (the latter being the value extrapolated from a Sellmeier fit of the experimental ellipsometric values shown in Fig. 1(b)). The devices under test are not single-mode around 2 µm. They can indeed sustain the propagation of at least the fundamental TE and TM modes. Owing to the rectangular symmetry of the waveguides, the properties of those two propagation modes differ significantly from one another, as illustrated in Fig. 2 for the case of a width of 1.8 µm, where effective indices of 1.937 and 1.60 are obtained for the two fundamental modes. Whereas the TE mode is highly confined within the SRN waveguide core (65%), the TM mode is much more delocalized (31%), so that a non-negligible part of the field propagates in the SiO$_2$ cladding. The difference in the field confinement is also reflected in the value of the effective area that is estimated as 0.86 µm$^2$ and 4.75 µm$^2$ for the TE and TM modes, respectively. Taking into account a nonlinear refractive index $n_2$ of 2.6 x 10$^{-20}$ m²/W and 1.8 x 10$^{-18}$ m²/W for SiO$_2$ and SRN respectively, this leads to a nonlinear Kerr coefficient $\gamma = 2\pi n_2 / \lambda A_{eff}$ of 6.56 W$^{-1}$.m$^{-1}$ and 1.2 W$^{-1}$.m$^{-1}$ for the TE and TM mode, respectively. Even though these nonlinear coefficients remain well below the record values that can be achieved with Si or more recently with AlGaAs [31], they are however more than one order of magnitude above the results achieved in stoichiometric Si$_3$N$_4$.

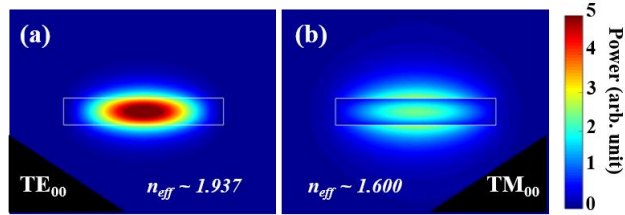

Fig. 2. Examples of the power distribution of the fundamental **(a)** TE-mode and **(b)** TM-mode. Results from numerical simulations of a waveguide having a width of 1.8 µm.

It is noted that dispersion is another important parameter of a wavelength converting nonlinear device, and should ideally be slightly anomalous at the wavelengths of interest. However, we have not attempted to optimize the design of the waveguides for efficient dispersion management, so that the dispersion is mainly imposed by the strongly normal dispersion of the material. At 2 µm, the dispersion coefficient is therefore evaluated to be around -593 ps/nm/km and -3210 ps/nm/km for the TE and TM modes respectively for a 1.8 µm wide waveguide. These dispersion values varied only moderately (within 5 %) in the range of waveguides we had available.

## 3. Experimental setup

Light coupling in the waveguide was achieved through butt-coupling assisted with lensed fibers. Due to the high birefringence of the waveguide, the input polarization will highly impact the coupling efficiency. Therefore, polarization controllers (PC) were used to excite the waveguide with either a TE or TM input field. The difference in loss between the two polarizations was found to be around 10 dB, with the TM mode exhibiting the higher losses. Subsequently, launching into the TE mode, we tested waveguides of different widths and found that the total fiber-to-fiber transmission (including coupling loss and linear propagation losses of the waveguide) was rather constant around 20 dB for widths between 1.6 and 2.2 µm.

In order to demonstrate the suitability of the SRN device for 2-µm optical communications, we have implemented the experimental setup depicted in Fig. 3(a), similar to the one used in [7] and based on commercially available 2-µm devices. The transmitter (TX) was based on an intensity modulated laser diode centered at 1980 nm, coupled to a Niobate-Lithium intensity modulator (IM). The Non-Return-to-Zero On-Off-Keying signal under test was a 2$^{31}$-1 pseudorandom bit sequence (PRBS) at 10 Gbit/s. Then a first thulium doped fiber amplifier (TDFA) was used to boost the signal before injection into the waveguide. The receiver (RX) was based on a second TDFA. This TDFA was set to work at constant gain. Therefore, a variable optical attenuator was employed at the TDFA output ensuring that the photodiode was operating at a constant power level. An optical bandpass filter (OBPF, 0.64 nm bandwidth) was also inserted at the output of the system in order to limit the accumulation of amplified spontaneous emission from the TDFAs. The optical signal was finally detected with a photodiode (PD) with an electrical bandwidth of 12.5 GHz. An optical spectrum analyzer (OSA) was used to evaluate the optical signal to noise ratio (OSNR) controlled thanks to the first variable optical attenuator.

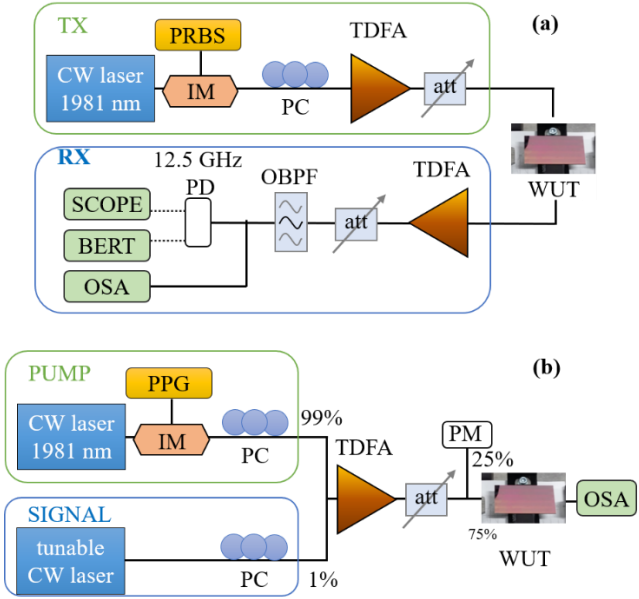

Fig. 3. Experimental setups used for (a) 10-Gbit/s transmissions, (b) wavelength conversion.

A second objective of our experiments was to evaluate the potential of the SRN waveguides for wavelength conversion applications. Details of the setup for this experiment are provided in Fig. 3(b). A CW laser tunable between 1965 and 1985 nm was used to generate a signal wave, while the pump beam was kept constant at 1981 nm. By means of an intensity modulator driven by an electrical pulse pattern generator (PPG), we were able to study the wavelength conversion process in pulsed pump regime with a typical pulse duration of 100 ps. The delay between two consecutive pulses could be adjusted from 200 ps to 1.6 ns, corresponding to repetition rates between 625 MHz and 5 GHz. The pump and signal waves were combined using a 99/1 coupler before being simultaneously amplified by a TDFA. Two polarization controllers were used to ensure that the pump and signal are launched into the TE mode. The optical power at the waveguide input was adjusted with a variable attenuator and monitored by a power meter. At the output of the system, an OSA was used to evaluate the conversion efficiency (CE), defined here as the ratio between the output powers of the signal and idler waves.

## 4. Data transmission at 2 µm

We have summarized the results obtained from the 10-Gbit/s transmission at 2 µm in Fig. 4. The transmission used the TE mode of propagation and the typical average power injected to the waveguide was of the order of 1 mW. Eye-diagrams are provided for the back-to-back measurement (panel (a)) as well as at the output of the waveguides with widths 1.6, 1.8 and 2 µm (panels (b)). In all cases, widely open eyes were obtained, indicating that the inclusion of the waveguides under test did not induce any major impairments on the transmission.

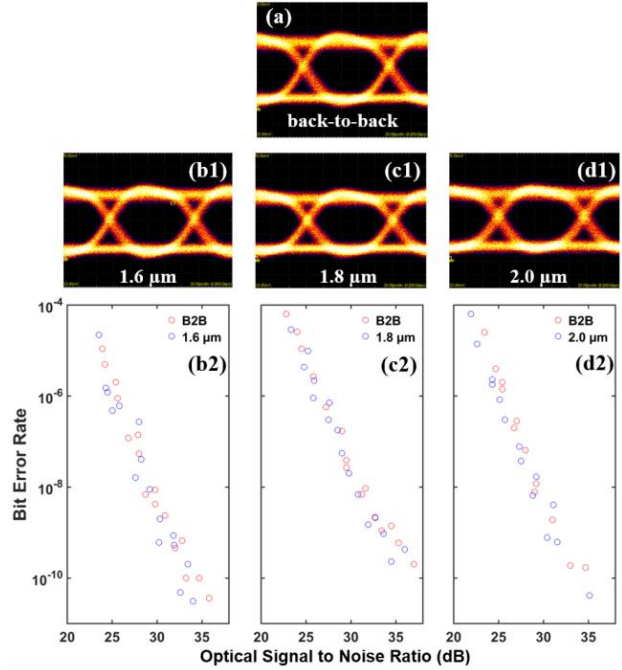

Fig. 4. (a) 10-Gbit/s eye diagram recorded in the back-to-back configuration compared to eye diagrams recorded after propagation in a 3-cm long waveguide (b). (c) Bit-error-rate measurements (BER) as a function of OSNR. Red circles are for back-to-back. Subplots correspond to results for the waveguides with a width of 1.6 µm (subplots 1), 1.8 µm (subplots 2) and 2 µm (subplots 3).

The quality of the 10-Gbit/s transmitted signal was quantitatively assessed through systematic measurements of the bit-error-rate as a function of the OSNR at the receiver. Results for the three waveguides under study are plotted in panels (c) and compared to the back-to-back (B2B) configuration. In all cases, BERs well below $10^{-9}$ were measured and no significant penalty was observed after transmission through the waveguide, confirming that SRN waveguides are perfectly suited for high-speed communications at 2 µm.

## 5. Wavelength conversion at 2 µm

We next study the wavelength conversion process. Three examples of wavelength conversion achieved for different wavelength detunings are shown in Fig. 5(a) for an estimated average input pump power of 15 dBm launched in the TE mode of the 1.8 µm wide waveguide. This corresponds to a peak power of the pulsed pump of almost 330 mW. Note that we have checked that the TM mode does not allow an efficient conversion.

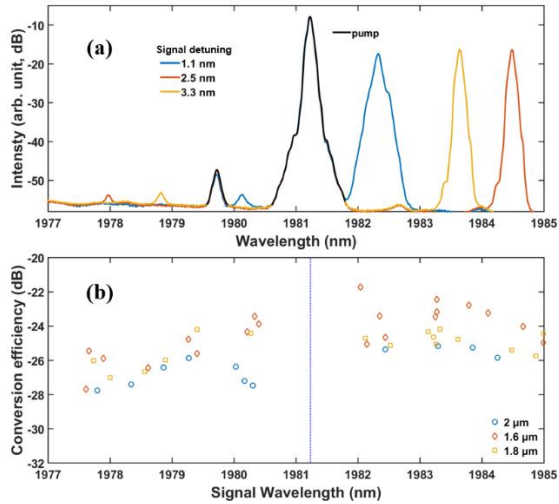

Fig. 5. (a) Wavelength conversion achieved in the 1.8 µm wide component for a signal detuned by 1.1, 2.5 and 3.3 nm (blue, yellow and red curves respectively). The black line is the pump spectrum. (b) Evolution of the conversion efficiency as a function of the signal wavelength. Results were recorded for 1.6, 1.8 and 2 µm width samples (red diamond, yellow square and blue circle respectively). Results achieved in pulsed regime with a peak power of 330 mW.

A more exhaustive study is reported in Fig. 5(b) showing an efficient conversion process of the idler wave detuned up to 3.3 nm from the pump. The moderate bandwidth achieved here denotes that, as expected, the pump lies in the strongly normal regime of dispersion. We also checked waveguides of 1.6 and 2.0 µm, which showed a rather similar level of performance, and conclude that the dispersion and nonlinear coefficients remain quite constant for this range of widths.

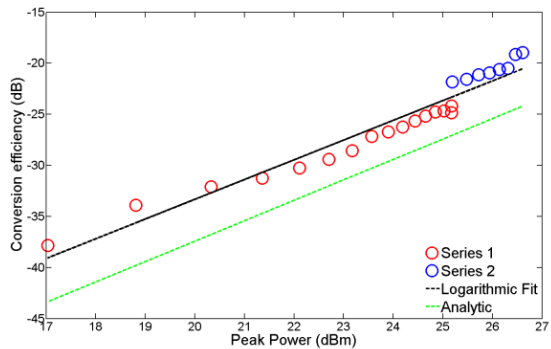

Fig. 6. Evolution of the conversion efficiency versus peak power of the pulsed signal for a converted wavelength of 1982.2 nm (red circles for a first series and blue circles for a second series), solid-line depicts for a logarithmic fit while the green dashed lines is for the analytic values. These results are obtained for the TE configuration launched in the 1.8 µm width waveguide.

Finally, we have also tested the conversion efficiency with respect to the input peak power of the pulsed pump. Results of two sets of measurements recorded for a signal wavelength of 1982.2 nm are summarized in Fig. 6 and show that conversion efficiencies close to -18 dB can be achieved for a peak-power reaching around 460 mW. The conversion efficiency of the degenerate four wave mixing increases monotonously with launched power. It scales as $P_P^{1.93}$ which is in agreement with the expected exponent 2. However, there is a 2-dB discrepancy on the peak power with the analytic model developed in Ref. [32]. This can be explained either by an underestimation of the coupled peak power since the propagation losses are not accurately known (estimated to 2dB/cm) or by an underestimation of the nonlinear index of the material at this wavelength. Indeed, at 2 µm, it is well known that the silicon nonlinear index increases, therefore the silicon rich silicon nitride might evolve similarly [33, 34].

## 6. Conclusion

To conclude, we demonstrated that subwavelength SRN waveguides can sustain error-free transmission of high-speed telecom signals around 2 µm on device lengths up to 3.5 cm without any significant penalty. We have also shown that the same nonlinear photonic chip is suitable for frequency conversion process with a conversion efficiency up to -18 dB when operating in pulsed regime and in the TE mode. Note that given the moderate level of maturity of the various 2-µm components that are available to date (e.g. too low extinction ratio for filters, amplifiers working only for input power higher than 1 mW, etc.), such characterization of telecommunication signals is not as straightforward as in the C-band. Consequently, it has not been possible to perform a wavelength conversion process on genuine encoded data and evaluate the quality of such a converted signal in the present study. However, both the error-free transmission experiments as well as the wavelength conversion study confirm that the 2-µm spectral window is a relevant and promising alternative deserving further investigations in a near future, especially in the context of high speed optical processing based on low-cost and ultra-compact optical chips combined with electronic circuits. A particular point that deserves further attention is the optimization of the component design. Indeed, using a thicker layer of SRN should be a way to overcome the detrimental strong normal dispersion [25] as well as to increase the power launched in the TE mode which presents the highest nonlinearity. With such technologically feasible improvements, wavelength conversion of data encoded signal under CW operation should be realistically demonstrated in a near future.

### Acknowledgments

This work is financially supported by PARI PHOTCOM Région Bourgogne, by Carnot Arts Institute (PICASSO 2.0 project), by the Institut Universitaire de France, by FEDER-FSE Bourgogne 2014/2020 and by the the french "Investissements d'Avenir" program, project ISITE-BFC (contract ANR-15-IDEX-0003). The research work has benefited from the PICASSO experimental platform of the University of Burgundy. The waveguides were fabricated in the clean-room complex of the Zepler Institute, University of Southampton.

### References

1. Garcia Gunning, F., et al., *Key Enabling Technologies for Optical Communications at 2000 nm.* Appl. Opt., 2018.
2. Zhang, H., et al., *100 Gbit/s WDM transmission at 2 µm: transmission studies in both low-loss hollow core photonic bandgap fiber and solid core fiber.* Opt. Express, 2015. **23**(4): p. 4946-4951.
3. Sadiq, M.U., et al., *10 Gb/s InP-based Mach-Zehnder modulator for operation at 2 µm wavelengths.* Opt. Express, 2015. **23**(9): p. 10905-10913.
4. Li, J., et al., *2µm Wavelength Grating Coupler, Bent Waveguide, and Tunable Microring on Silicon Photonic MPW.* IEEE Photon. Technol. Lett., 2018. **30**(5): p. 471-474.
5. Lamy, M., et al., *Ten gigabit per second optical transmissions at 1.98 µm in centimetre-long SiGe waveguides.* Electron. Lett., 2017. **53**(17): p. 1213-1214.
6. Lamy, M., et al., *10 Gbps data transmission in TiO2 waveguides at 2 um.* Appl. Sci., 2017. **7**: p. 631.
7. Lamy, M., et al., *Silicon Waveguides for High-Speed Optical Transmissions and Parametric Conversion Around 2 µm.* IEEE Photon. Technol. Lett., 2019. **31**(2): p. 165-168.
8. Philipp, H.T., et al., *Amorphous silicon rich silicon nitride optical waveguides for high density integrated optics.* Electron. Lett., 2004. **40**(7): p. 419-421.
9. Rahim, A., et al., *Expanding the Silicon Photonics Portfolio With Silicon Nitride Photonic Integrated Circuits.* J. Lightw. Technol., 2017. **35**(4): p. 639-649.

10. Ettabib, M.A., et al., *Highly nonlinear tellurite glass fiber for broadband applications,"* in *Optical Fiber Communication Conference* 2014: San Francisco, CA. p. Tu2K.3.
11. Xing, S., et al., *Mid-infrared continuous-wave parametric amplification in chalcogenide microstructured fibers.* Optica, 2017. **4**(6): p. 643-648.
12. Cheng, T., et al., *Broadband cascaded four-wave mixing and supercontinuum generation in a tellurite microstructured optical fiber pumped at 2 μm.* Opt. Express, 2015. **23**(4): p. 4125-4134.
13. Borghi, M., et al., *Nonlinear silicon photonics.* J. Opt, 2017. **19**(9): p. 093002.
14. Lacava, C., A.M. Ettabib, and P. Petropoulos, *Nonlinear Silicon Photonic Signal Processing Devices for Future Optical Networks.* Appl. Sci., 2017. **7**(1).
15. Zlatanovic, S., et al., *Mid-infrared wavelength conversion in silicon waveguides using ultracompact telecom-band-derived pump source.* Nat. Photon., 2010. **4**: p. 561.
16. Kuyken, B., et al., *50 dB parametric on-chip gain in silicon photonic wires.* Opt. Lett., 2011. **36**(22): p. 4401-4403.
17. Liu, X., et al., *Mid-infrared optical parametric amplifier using silicon nanophotonic waveguides.* Nat. Photon., 2010. **4**: p. 557.
18. Lau, R.K.W., et al., *Continuous-wave mid-infrared frequency conversion in silicon nanowaveguides.* Opt. Lett., 2011. **36**(7): p. 1263-1265.
19. Lamy, M., et al., *Broadband embedded metal grating couplers in Titanium dioxide waveguides.* Opt. Lett, 2016. **42**(14): p. 2778-2781.
20. Krückel, C.J., et al., *Continuous wave-pumped wavelength conversion in low-loss silicon nitride waveguides.* Opt. Lett., 2015. **40**(6): p. 875-878.
21. Moss, D.J., et al., *New CMOS-compatible platforms based on silicon nitride and Hydex for nonlinear optics.* Nat. Photon., 2013. **7**: p. 597.
22. Tan, D.T.H., K.J.A. Ooi, and D.K.T. Ng, *Nonlinear optics on silicon-rich nitride - a high nonlinear figure of merit CMOS platform.* Photonics Research, 2018. **6**(5): p. B50-B66.
23. Krückel, C.J., et al., *Linear and nonlinear characterization of low-stress high-confinement silicon-rich nitride waveguides.* Opt. Express, 2015. **23**(20): p. 25827-25837.
24. Dizaji, M.R., et al., *Silicon-rich nitride waveguides for ultra-broadband nonlinear signal processing.* Opt. Express, 2017. **25**(11): p. 12100-12108.
25. Lacava, C., et al. *All-optical Wavelength Conversion of Phase-encoded Signals in Silicon-rich Silicon Nitride Waveguides*. in *Conference on Lasers and Electro-Optics*. 2018. San Jose, California: Optical Society of America.
26. Ooi, K.J.A., et al., *Pushing the limits of CMOS optical parametric amplifiers with USRN:Si7N3 above the two-photon absorption edge.* Nat Commun, 2017. **8**: p. 13878.
27. Wang, T., et al., *Supercontinuum generation in bandgap engineered, back-end CMOS compatible silicon rich nitride waveguides.* Laser Photonics Rev., 2015. **9**(5): p. 498-506.
28. Choi, J.W., et al., *Wideband nonlinear spectral broadening in ultra-short ultra - silicon rich nitride waveguides.* Sci. Rep., 2016. **6**: p. 27120.
29. Lacava, C., et al., *Si-rich Silicon Nitride for Nonlinear Signal Processing Applications.* Sci. Rep., 2017. **7**(1): p. 22.
30. Dominguez Bucio, T., et al., *Material and optical properties of low-temperature NH$_3$-free PECVD SiN$_x$ layers for photonic applications.* J. Phys. D: Appl. Phys., 2017. **50**(2): p. 025106.
31. Pu, M., et al., *Ultra-Efficient and Broadband Nonlinear AlGaAs-on-Insulator Chip for Low-Power Optical Signal Processing.* Laser Photonics Rev., 2018. **0**(0): p. 1800111.
32. Ebnali-Heidari, M., et al., *A proposal for enhancing four-wave mixing in slow light engineered photonic crystal waveguides and its application to optical regeneration.* Opt. Express, 2009. **17**(20): p. 18340-18353.
33. Hon, N.K., R. Soref, and B. Jalali, *The third-order nonlinear optical coefficients of Si, Ge, and Si$_{1−x}$Ge$_x$ in the midwave and longwave infrared.* J. Appl. Phys., 2011. **110**(1): p. 011301.
34. Bristow, A.D., N. Rotenberg, and H.M. van Driel, *Two-photon absorption and Kerr coefficients of silicon for 850–2200nm.* Appl. Phys.Lett., 2007. **90**(19): p. 191104.